\documentclass{article}
\usepackage{cmap}
\usepackage{arxiv}
\usepackage[T1]{fontenc}
\usepackage{amsmath,amssymb}
\usepackage{graphicx}
\usepackage{float}
\graphicspath{{figures/}}
\usepackage{booktabs}
\usepackage{array,tabularx}
\usepackage{xltabular}
\usepackage{xcolor}
\usepackage{microtype}
\usepackage{etoolbox}
\usepackage{placeins}
\usepackage{needspace}
\usepackage{url}
\usepackage[numbers,sort&compress]{natbib}
\usepackage[colorlinks=true,linkcolor=blue,citecolor=blue,urlcolor=blue]{hyperref}
\usepackage[nameinlink,noabbrev]{cleveref}

\newcommand{\bench}{ICLayout-Bench}
\newcommand{\benchTaskCount}{165}
\newcommand{\benchPdkCount}{13}
\newcommand{\benchPublicCount}{92}
\newcommand{\benchPrivateCount}{73}
\newcommand{\benchMinNode}{3}
\newcommand{\benchMaxNode}{180}
\newcolumntype{Y}{>{\raggedright\arraybackslash}X}
\newcolumntype{P}[1]{>{\raggedright\arraybackslash}p{#1}}
\title{ICLayout-Bench: Benchmarking LLM Agents for\\
Custom Circuit Layout Design}
\author{%
  \textbf{Zhangmai Li\textsuperscript{1,2} \quad
  Tianrui Ma\textsuperscript{2,*} \quad
  Zheng Weng\textsuperscript{2,3} \quad
  Mo Zou\textsuperscript{2} \quad
  Zidong Du\textsuperscript{2,3}}\\[3pt]
  \textbf{Yifan Hao\textsuperscript{2,3} \quad
  Yongwei Zhao\textsuperscript{2,3} \quad
  Wei Li\textsuperscript{2,3} \quad
  Jing Li\textsuperscript{1} \quad
  Qi Guo\textsuperscript{2,3} \quad
  Yunji Chen\textsuperscript{2,4,3}}\\[6pt]
  \textnormal{\small\textsuperscript{1}University of Science and Technology of China}\\
  \textnormal{\small\textsuperscript{2}State Key Lab of Processors, Institute of Computing Technology, Chinese Academy of Sciences}\\
  \textnormal{\small\textsuperscript{3}University of Chinese Academy of Sciences}\\
  \textnormal{\small\textsuperscript{4}Institute of AI for Industries, Chinese Academy of Sciences}\\[4pt]
  \textnormal{\small\textsuperscript{*}Corresponding author:
  \href{mailto:matianrui@ict.ac.cn}{\nolinkurl{matianrui@ict.ac.cn}}}
}
\date{}
\renewcommand{\shorttitle}{ICLayout-Bench}
\renewcommand{\headeright}{}
\renewcommand{\undertitle}{}
\makeatletter
\patchcmd{\@maketitle}{\textsc{\undertitle}\\}{}{}{}
\patchcmd{\@maketitle}
  {\begin{tabular}[t]{c}\bf\rule{\z@}{24\p@}\@author\end{tabular}}
  {\ifx\@author\@empty\else
    \begin{tabular}[t]{c}\bf\rule{\z@}{24\p@}\@author\end{tabular}\fi}{}{}
\patchcmd{\@maketitle}
  {\vskip 0.4in \@minus 0.1in \center{\@date}   \vskip 0.2in}
  {\ifx\@date\@empty\vskip 0.08in\else
    \vskip 0.4in \@minus 0.1in \center{\@date}\vskip 0.2in\fi}{}{}
\makeatother
\hypersetup{
  pdftitle={ICLayout-Bench: Benchmarking LLM Agents for Custom Circuit Layout Design},
  pdfauthor={Zhangmai Li, Tianrui Ma, Zheng Weng, Mo Zou, Zidong Du, Yifan Hao, Yongwei Zhao, Wei Li, Jing Li, Qi Guo, Yunji Chen},
  pdfsubject={Custom circuit layout agent benchmarking},
  pdfkeywords={Custom circuit layout, LLM agents, Electronic design automation}
}

\begin{document}
\raggedbottom
\maketitle

\begin{abstract}
Large language model (LLM) agents are being explored for circuit design,
but their ability to produce high-quality custom layouts has not been
extensively evaluated.
\bench{} is a benchmark in which agents use electronic design
automation (EDA) tools to construct and optimize GDSII layouts from
transistor-level circuit netlists.
These layouts are evaluated using physical verification, parasitic
extraction, and post-layout simulation. Layouts that pass the physical and
functional checks are scored on electrical performance and area.
The benchmark includes \benchTaskCount{} tasks across
\benchPdkCount{} process design kits (PDKs), covering technology nodes from
\benchMinNode{}\,nm to \benchMaxNode{}\,nm.
The tasks span analog, mixed-signal, and digital circuits in
open-source and commercial EDA flows.
The experiments evaluate 22 configurations spanning seven models on
21 public and 19 private tasks. Mean layout scores range from 4.22 to
80.26 on public tasks and from 1.43 to 67.01 on private tasks, on a
100-point scale. Higher reasoning effort does not consistently improve
mean scores, and configurations with comparable scores can incur
substantially different model costs.
\par\smallskip
\noindent Project Page: \url{https://lizhangmai.github.io/ICLayout-Bench}\\
Repository: \url{https://github.com/lizhangmai/ICLayout-Bench}\\
Dataset: \url{https://huggingface.co/datasets/lizhangmai/ICLayout-Bench}
\end{abstract}

\keywords{Custom circuit layout \and LLM agents
\and Electronic design automation}

\section{Introduction}
\label{sec:introduction}

Large language models (LLMs) are increasingly used for circuit design,
from generating circuit descriptions \citep{rtlcoder2025,analogcoder2025}
to revising designs through tool-driven iteration \citep{autochip2024}.
Existing benchmarks assess the functional
correctness and implementation quality of generated
register-transfer-level (RTL) designs
\citep{rtllm2024,verilogeval2023,pluto2025}, and the structural
correctness and electrical performance of transistor-level circuit
descriptions \citep{analogdesignbench2026,sganalog2026}.
Layout-oriented benchmarks also test understanding of existing geometry
\citep{chiplayoutbench}.
However, existing benchmarks have not fully assessed agents' ability
to construct valid custom layouts and optimize their post-layout
electrical performance and area.

Custom layout realizes a circuit's devices and connections as geometry
on process layers. Placement and routing affect parasitic resistance
and capacitance and hence circuit response, while the geometry also
determines area \citep{align2020,icpi2026}.
Physical verification checks process-rule compliance and circuit
consistency, while simulation after parasitic extraction checks
functional requirements such as correct output polarity, logic
transitions, and state retention. Even layouts that satisfy all these
checks can differ in electrical performance and area.
Evaluation must therefore check physical and functional validity before
assessing post-layout electrical performance and area efficiency.

\begin{figure}[t]
  \centering
  \includegraphics[width=\linewidth]{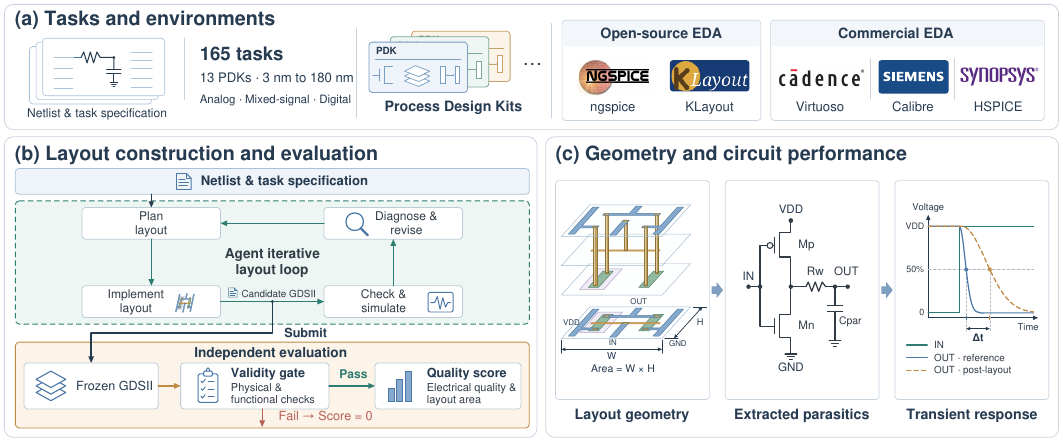}
  \caption{Overview of \bench{}: (a) task inputs, process coverage, and EDA environments;
  (b) iterative layout construction and independent evaluation of frozen
  GDSII submissions; (c) an illustrative CMOS inverter example linking
  layout geometry, parasitic resistance and capacitance, and transient
  response.}
  \label{fig:overview}
\end{figure}

\bench{} is a benchmark for constructing and optimizing custom
layouts from complete transistor-level circuits.
Each task fixes the circuit, permitted resources, implementation
constraints, and electrical tests. Agents use electronic design
automation (EDA) tools and their feedback to construct and revise layouts.
An independent evaluator reruns physical and functional checks on frozen
GDSII submissions and scores valid layouts on electrical performance
and area.
The suite contains \benchTaskCount{} tasks spanning \benchPdkCount{}
process design kits (PDKs), covering technology nodes from
\benchMinNode{}\,nm to \benchMaxNode{}\,nm.
Its public and private collections cover analog,
mixed-signal, and digital circuits in open-source and commercial
EDA flows.
\Cref{fig:overview} summarizes the task coverage and iterative workflow.

The experiments evaluate a total of 22 configurations spanning seven
models on 21 public and 19 private tasks. Each configuration combines
a model, harness, and reasoning-effort setting.
On these subsets, mean layout scores do not consistently increase
with reasoning effort. Configurations with similar mean scores
can also differ substantially in model cost.
Alongside layout scores, we report model cost and the
token-to-score ratio ($R_T$), which measures output tokens per score point.

\section{ICLayout-Bench}
\label{sec:construction}

\subsection{Task Definition and Solver Interface}
\label{sec:task-interface}

Each task supplies a transistor-level circuit netlist, external ports,
permitted PDK and EDA resources, implementation constraints, and electrical
tests. The circuit fixes device types, dimensions,
multiplicities, body connections, and connectivity. The agent must
realize that circuit as a GDSII layout containing the specified top cell,
choosing placement, routing, hierarchy, fingering, and diffusion sharing
within the declared constraints.

Tasks share a common input and submission format. Each task fixes its
PDK, process models, and EDA toolchain. The evaluation plan specifies
rule decks for physical verification, settings for parasitic extraction,
and operating conditions for electrical tests.

\bench{} exposes a session-scoped Model Context Protocol (MCP) bridge
that connects native agent harnesses to the EDA service. The bridge
provides a common interface for workspace file access, command execution,
monitoring long-running jobs, report retrieval, and candidate submission.
Its \texttt{check} operation runs the task's full evaluation plan and
returns physical and functional diagnostics and raw electrical
measurements.
The last GDSII snapshot frozen by the \texttt{submit} operation is
evaluated independently outside the writable solver workspace.

\subsection{Task Construction and Coverage}
\label{sec:task-coverage}

Task construction begins with a complete circuit and a compatible PDK,
with the external interface and physical layout constraints specified.
Electrical tests are designed around the circuit's intended function and
the effects captured by the device and extraction models. The tests
specify stimuli, loads, and operating conditions. Functional requirements
are defined separately from quality objectives for electrical performance
and area. Engineering targets or pre-layout measurements provide the
scoring baselines, while weights reflect the task objectives and each
metric's sensitivity to layout under the declared models.

\begin{figure}[t]
  \centering
  \includegraphics[width=0.65\linewidth]{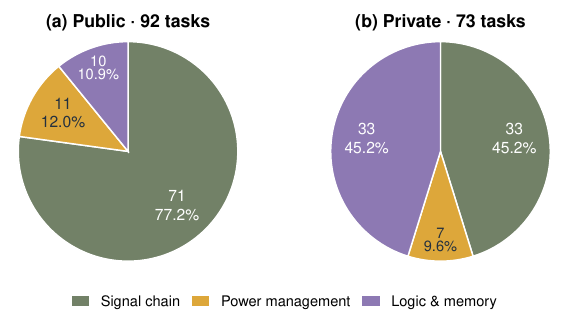}
  \caption{Circuit-group composition of the full public (a) and private (b)
  inventories. Slice labels give task counts and percentages.}
  \label{fig:task-coverage}
\end{figure}

The inventory contains \benchTaskCount{} tasks, divided into
\benchPublicCount{} public and \benchPrivateCount{} private tasks.
Public tasks are adapted from published circuit repositories, including
\href{https://github.com/ferdous313/OpenRAM}{OpenRAM}~\citep{openram2016},
\href{https://github.com/MacAnalog/spicexplorer-release}{analog-db}~\citep{analogdb2026},
\href{https://github.com/idea-fasoc/OpenFASOC}{OpenFASOC/GLayout}~\citep{openfasoc2023,glayout2024}, and
\href{https://github.com/efabless/sky130_ef_ip__ccomp3v}{Efabless} designs~\citep{efablesscomparator};
private tasks come from tapeout-validated designs.
The suite covers signal-chain circuits, power management circuits,
and logic and memory circuits. The public collection is
concentrated in signal-chain circuits, while the private collection
has similar numbers of signal-chain and logic and memory tasks.
\Cref{fig:task-coverage} compares the circuit-group proportions in
the two full inventories.

These circuits span \benchPdkCount{} PDKs at technology nodes from
\benchMinNode{}\,nm to \benchMaxNode{}\,nm.
Public tasks use KLayout~\citep{klayout}, Magic~\citep{magic1984},
Netgen, KPEX, and ngspice~\citep{ngspice}.
Private flows use Virtuoso and Custom Compiler for layout construction;
Calibre, IC Validator, PVS, and Assura for physical verification;
Calibre PEX, StarRC, and Quantus for parasitic extraction; and Spectre
and HSPICE for simulation.
\Cref{tab:inventory} gives the detailed process coverage.

\subsection{Independent Evaluation and Scoring}
\label{sec:evaluation}
\label{sec:scoring}
\label{sec:outcomes}

Final evaluation independently reruns the checks on a frozen GDSII
submission using the circuit, rule decks, and test conditions specified
by the task. It first verifies the declared cell, ports, and physical
constraints. Design-rule checking (DRC) applies
the task's rule deck, and layout-versus-schematic (LVS) comparison checks
device parameters and connectivity, including body connections.
After parasitic extraction, the evaluator simulates the circuit with
the extracted parasitics, using the task's models, loads, stimuli, and
operating conditions.
The simulations test required behavior, such as output polarity, logic
transitions, or retained state, and measure electrical performance for
quality scoring.
Compared configurations use the same rule decks, device models, tests,
and extraction settings on a given task. Extraction coverage can differ
across tasks, including capacitance-only and distributed-RC models.
Each result includes the extraction and toolchain settings needed to
interpret its electrical measurements.

A submission is valid ($G=1$) when all required physical and functional
checks pass. A failed check sets $G=0$ and the score to zero. Attempts
without a submitted layout also receive zero.

For each electrical objective, the normalized credit $c_j$ is calculated
using a predefined function and scale. Its anchor is a declared
engineering target or, if none is specified, a pre-layout measurement of
the supplied circuit under the same conditions. Each required condition
is normalized separately; the lowest normalized value is retained and
capped at one. \Cref{app:interface} gives the normalization functions and
condition aggregation. The area credit is
$c_A=\min(1,A_{\mathrm{target}}/A)$, where $A$ is the bounding-box area
over the declared functional layers.

Using the predefined weights $w_j$, the score combines these credits
through a geometric mean and applies the validity gate:
\begin{equation}
  S=100\,G\prod_{j\in\mathcal J}c_j^{w_j},
  \qquad 0\leq c_j\leq1,\quad
  \sum_{j\in\mathcal J}w_j=1,\quad w_j>0,
  \label{eq:score}
\end{equation}
where $\mathcal J$ contains the positively weighted electrical and area
objectives, including at least one electrical objective. Because each
credit is capped, exceeding one target cannot offset a shortfall in
another objective. A valid layout receives 100 only when every weighted
credit equals one. The result retains raw measurements and component
credits alongside the aggregate score.

\section{Experiments}
\label{sec:experiments}

\subsection{Experimental Setup}
\label{sec:experimental-setup}

The experiments reported in this paper use fixed subsets of
21 public and 19 private tasks from the
inventory in \Cref{sec:task-coverage}.
To limit evaluation time, we select representative tasks
with broad coverage of circuit types, PDKs, and EDA toolchains while
reducing redundancy.
Both subsets cover all three circuit groups. Their detailed composition
appears in \Cref{tab:inventory}.
The seven evaluated models are GPT-6 Astra, GPT-6.1 Sol,
GPT-6.1 Terra, GPT-6 Luna, DeepSeek Flash (V4.1 Flash, 1017 release), Kimi K3.1 Pro,
and GLM-5.3 Flash.
The models use their providers' official native harnesses: Codex for the
GPT models, DeepSeek Harness (DSH) for DeepSeek Flash, Kimi Code for Kimi K3.1 Pro, and
ZCode for GLM-5.3 Flash. Evaluating different reasoning-effort settings yields a
total of 22 configurations per collection.
Every configuration is assigned all
selected tasks in its collection. Effort labels are provider-specific,
so identical labels do not imply equivalent reasoning effort across models.

On each task, compared configurations receive the same circuit,
permitted process resources, constraints, tests, and time budget.
Task-specific wall-clock budgets range from 1 to
12 hours and include checks and EDA operations.
Reference layouts and task-specific implementation code are excluded
from solver inputs.

\subsection{Layout Scores}
\label{sec:main-results}

\begin{figure}[H]
  \centering
  \includegraphics[width=0.90\linewidth]{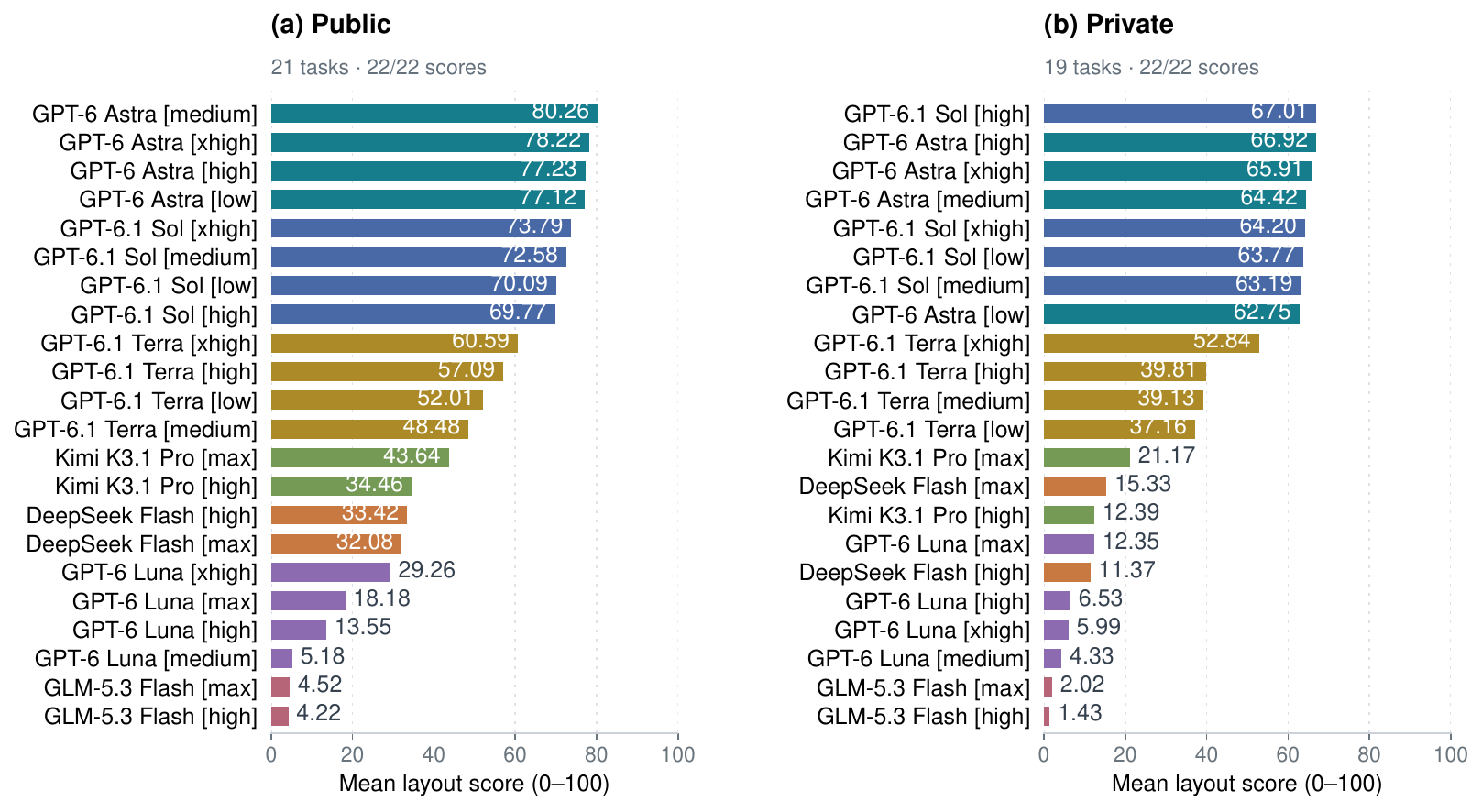}
  \caption{Mean layout scores on the public (a) and private (b)
  experimental subsets. Rows represent configurations,
  with effort settings in brackets, and are ranked by mean score.
  Colors identify models.}
  \label{fig:main-results}
\end{figure}

Mean layout scores are unweighted averages over all selected tasks,
including zero scores for failed attempts and attempts without a submitted
layout (\Cref{fig:main-results}). Public mean scores range from
4.22 to 80.26, led by
GPT-6 Astra [medium]; private means range from 1.43
to 67.01, led by GPT-6.1 Sol [high].
The per-task breakdown appears in \Cref{app:task-scores}.
In these experiments, higher reasoning effort does not consistently
improve mean scores (\Cref{fig:main-results}).
For example, GPT-6 Astra's public mean score decreases from 80.26 at medium
effort to 77.23 at high effort.

\subsection{Model Cost and Token-to-Score Ratio}
\label{sec:efficiency}

\Cref{fig:resource-efficiency} plots mean layout scores against model
cost per task and the token-to-score ratio for the same task set.

\begingroup
\setlength{\intextsep}{3pt}
\setlength{\abovecaptionskip}{4pt}
\begin{figure}[H]
  \centering
  \includegraphics[width=0.98\linewidth,trim=0 5pt 0 5pt,clip]{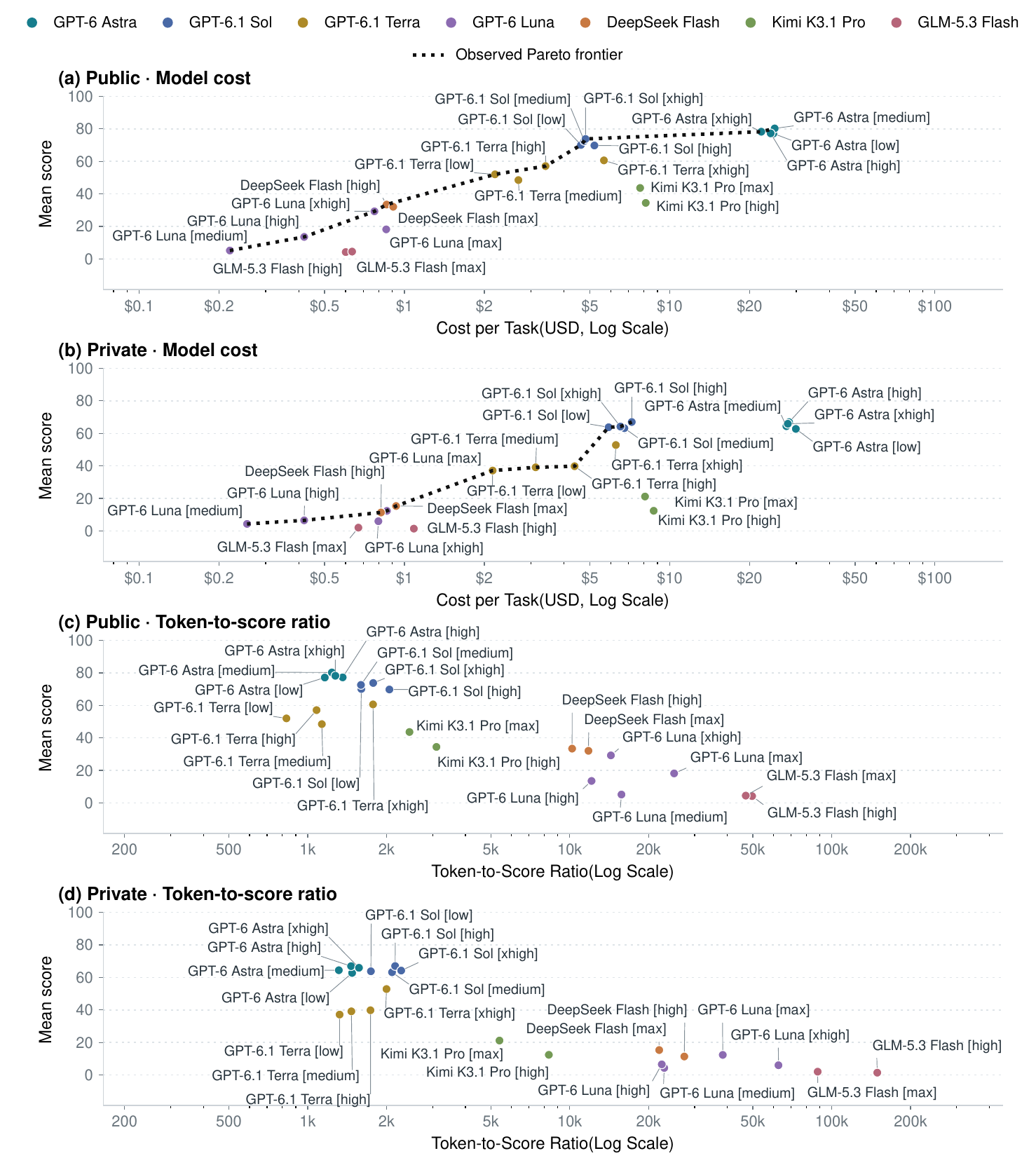}
  \caption{Mean score versus model cost per task (a, b) and the
  token-to-score ratio (c, d) for public (a, c) and private (b, d) tasks.
  Dotted lines mark the cost--score Pareto frontiers.}
  \label{fig:resource-efficiency}
\end{figure}
\endgroup

Model cost per task is computed from
recorded usage and declared USD pricing, including input, output, and
cache accounting where available. Unpriced compute and EDA licenses
are excluded. Output-token counts are provider-reported and include
reasoning tokens; tokenization conventions can differ across models.
The token-to-score ratio $R_T$ is defined as total output tokens divided
by total score:
\[
  R_T=\frac{\sum_{i\in\mathcal C}T_i}{\sum_{i\in\mathcal C}S_i}
     =\frac{\overline{T}}{\overline{S}},
\]
where $\mathcal C$ is the evaluated task set for the collection, $T_i$ is the
recorded output-token count for task $i$, and $S_i$ is its 0--100 score.
Lower $R_T$ means fewer output tokens for a given score.

On the private tasks, GPT-6 Astra [high] and GPT-6.1 Sol [high] achieve nearly equal
mean scores of 66.92 and 67.01. GPT-6 Astra has a lower token-to-score ratio
(1,464 versus 2,157 output tokens per score point), while its model cost
is higher (\$28.19 versus \$7.19 per task).

For GPT-6 Astra and GPT-6.1 Sol at medium effort on public tasks and high effort on
private tasks, cached-input charges account for
57.8--73.0\% of model cost, compared with 17.4--27.2\% for output generation.

\begin{figure}[H]
  \centering
  \includegraphics[width=0.97\linewidth]{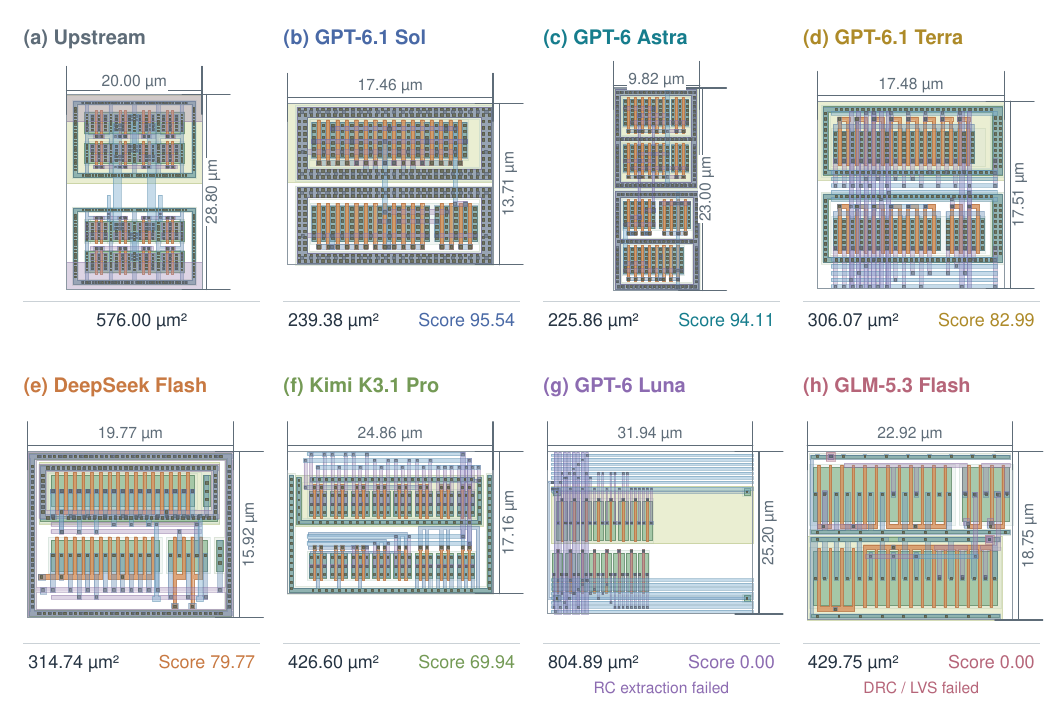}
  \caption{GF180 five-transistor OTA layouts. Panel (a) shows the unscored
  upstream layout \citep{otasource}; the other panels show model submissions
  at high effort, ordered by descending score from left to right and top to bottom.
  Dimensions and areas describe the GDS bounding boxes.
  Layouts are scaled independently to fit their panels while preserving aspect ratios.
  Failure labels identify the rejected evaluation stage.}
  \label{fig:case-study}
\end{figure}

\subsection{Case Study: Validity and Layout Quality}
\label{sec:case-studies}

The public GF180 five-transistor OTA task fixes the complete netlist,
including bias devices and dummy transistors \citep{otasource}, and gives
each independent attempt one hour to construct and optimize its layout.
Ordinary electrical tests use extracted distributed interconnect RC,
10\,$\mu$A bias, and nominal loads of 0.1, 1, and 5\,pF, with additional
process, voltage, and temperature conditions.
Area has 25\% weight against a 240\,$\mu\mathrm{m}^2$ target;
unity-gain bandwidth and phase margin carry 18\% and 7\%.
The transient objectives measure 0.1\% settling after a 20\,mV follower
step (15\%), slew rate after a 0.5\,V step (8\%), and sinusoidal
linearity (2\%). DC gain, input offset, input common-mode range, and
output swing together carry 7\%; common-mode and supply rejection
and integrated input noise carry 5\%; quiescent power carries 3\%.
Gain, offset, and intrinsic noise each retain 1\% weight because their
layout sensitivity is limited under the fixed circuit and deterministic
extraction model.

The remaining 10\% evaluates substrate and well immunity, with 5\% per
domain. Each test injects a zero-DC, 100\,$\mu$A AC current and measures
the follower's output disturbance at 13 frequencies from 1\,kHz to
1\,MHz, retaining the maximum over frequencies and four injection
directions. The declared external domains and return conditions are
fixed. A projected-sheet model resolves body contacts and supply
return paths with ideal signal interconnect; ordinary performance
tests use the separate distributed-RC extraction.
A 10\,$\mu$V disturbance gives half credit in either domain.
These tests assess immunity within the declared model, without
implying calibrated three-dimensional silicon isolation or mismatch yield.

\Cref{fig:case-study} compares the seven high-effort submissions with
the unscored upstream layout. Five submissions pass the complete
evaluation, scoring 69.94--95.54. GLM-5.3 Flash fails DRC and LVS; GPT-6 Luna passes
both but fails parasitic extraction.
GPT-6.1 Sol and GPT-6 Astra occupy 239.38 and 225.86\,$\mu\mathrm{m}^2$, respectively,
so both receive full area credit. GPT-6.1 Sol nevertheless scores 95.54 versus
GPT-6 Astra's 94.11 because its aggregate electrical credit $E$
(\Cref{app:interface}) is higher: 0.9410 versus 0.9223.
Their worst well-induced output disturbances are 0.72 and
2.61\,$\mu$V, respectively. Reducing area below the target gives no
additional credit to compensate for weaker electrical performance.

\Cref{fig:ota-effort} compares GPT-6.1 Terra's medium, high, and xhigh
submissions. The medium attempt moved horizontal buses over the device
banks, shared diffusion between adjacent fingers with a common terminal,
and tightened edge spacing to reach 239.73\,$\mu\mathrm{m}^2$.
In the high attempt, adding body-contact strips and side contacts reduced
the worst well-induced output disturbance from 1.31\,mV to
37.23\,$\mu$V while increasing area from 325.36 to
346.10\,$\mu\mathrm{m}^2$; subsequent compaction reduced area to
306.07\,$\mu\mathrm{m}^2$.
In the xhigh attempt, the PMOS mirror fingers were interleaved and the
dummy fingers redistributed to both row ends. At an unchanged area of
335.21\,$\mu\mathrm{m}^2$, the worst well-induced output disturbance
fell from 31.57 to 6.65\,$\mu$V under the declared model.
Further input-pair rearrangement and routing compaction produced the
final 306.24\,$\mu\mathrm{m}^2$ layout.

\begingroup
\setlength{\intextsep}{3pt}
\setlength{\abovecaptionskip}{4pt}
\begin{figure}[H]
  \centering
  \includegraphics[width=0.97\linewidth]{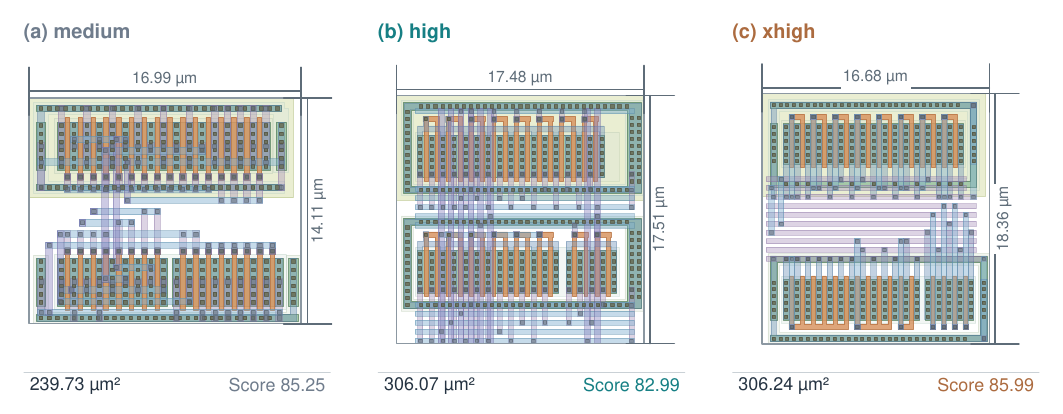}
  \caption{GPT-6.1 Terra OTA submissions at (a) medium, (b) high,
  and (c) xhigh effort. Layouts are scaled independently to fit
  their panels while preserving aspect ratios.
  All three pass the required physical and functional
  checks. Dimensions are marked on each layout; the labels below give
  bounding-box area and score.}
  \label{fig:ota-effort}
\end{figure}
\endgroup

\section{Related Work}
\label{sec:related}

\paragraph{LLM-assisted circuit design.}
ChipNeMo adapts LLMs to chip-design assistance and EDA script generation
\citep{chipnemo2023}. RTLCoder targets open-source RTL generation
\citep{rtlcoder2025}, while AutoChip iterates on HDL using compiler and
simulator feedback \citep{autochip2024}.
AnalogCoder constructs analog circuits through feedback-guided Python
code generation \citep{analogcoder2025}.
These methods study code and circuit generation; \bench{} evaluates
the physical implementation of a supplied circuit.

\paragraph{Circuit and layout benchmarks.}
AnalogGym standardizes simulation-based evaluation of analog sizing
algorithms across circuit families, PDKs, and simulators
\citep{analoggym2025}.
Analog Design Bench evaluates circuit design
against electrical specifications using an isolated verifier
\citep{analogdesignbench2026}. SGAnalog tests schematic transcription
through structural comparison and device sizing through
reference-relative electrical scores \citep{sganalog2026}.
These benchmarks evaluate circuit descriptions; constructing device-level
geometry falls outside their task scope.
ChipLayoutBench measures reasoning and grounding over existing layout
representations \citep{chiplayoutbench}.
\Cref{tab:positioning} compares the tasks, outputs, and evaluation
criteria of representative circuit and layout benchmarks.

\paragraph{Automatic layout generation.}
LAYGEN II combines reusable layout templates with evolutionary placement
and routing \citep{laygen2013}.
BAG supports parameterized circuit generators
\citep{bag2014}, while ALIGN and MAGICAL automate analog layout from
circuit descriptions \citep{align2020,magical2019}.
AutoCkt combines reinforcement-learning-based circuit sizing with
BAG-generated layouts and post-layout simulation \citep{autockt2020}.
GLayout supports language-guided generation through a technology-generic
representation; its benchmark reports compilation, DRC, and LVS
outcomes, with no post-layout electrical or area scores
\citep{glayout2024}.
PANDA coordinates topology, sizing, and layout
\citep{panda2026}.
For standard-cell layout generation, BonnCell combines routability-aware
transistor placement with routing based on mixed-integer programming
\citep{vancleeff2020bonncell}. SP\&R jointly optimizes transistor
placement and in-cell routing using satisfiability modulo theories
\citep{lee2021spr}, while NVCell uses reinforcement learning for
placement and design-rule repair \citep{ren2021nvcell}.
For agent-assisted refinement, LayoutCopilot translates user instructions
into edits to existing layouts \citep{layoutcopilot2024}, and ICPI
adjusts generator parameters using post-layout simulation
\citep{icpi2026}.

\begin{table}[H]
  \caption{Comparison of representative circuit and layout benchmarks
  by task, output, and evaluation.}
  \label{tab:positioning}
  \centering\small
  \setlength{\tabcolsep}{6pt}
  \renewcommand{\arraystretch}{1.22}
  \begin{tabularx}{\linewidth}{@{}P{3.6cm}P{3.1cm}P{2.2cm}Y@{}}
    \toprule
    Benchmark & Task & Output & Evaluation \\
    \midrule
    Analog Design Bench \citep{analogdesignbench2026}
    & Circuit design & SPICE netlist
    & Electrical specification compliance \\
    \addlinespace[3pt]
    SGAnalog \citep{sganalog2026}
    & Transcription / sizing & SPICE netlist
    & Circuit structure and performance \\
    \addlinespace[3pt]
    ChipLayoutBench \citep{chiplayoutbench}
    & Layout understanding & Answers
    & Geometric and spatial reasoning \\
    \midrule
    \textbf{\bench{}}
    & Layout construction & GDSII layout
    & Layout validity, performance, and area \\
    \bottomrule
  \end{tabularx}
\end{table}

\section{Discussion and Conclusion}
\label{sec:discussion}

\bench{} evaluates whether agents can combine circuit knowledge,
geometric reasoning, and EDA feedback to adapt their layout decisions
across circuits, processes, and EDA tools.
The experiments show that higher reasoning effort does not consistently
improve scores.
Model costs can differ substantially even when average scores are similar.
Future work will expand the task collection across technology nodes and
circuit types. The focus will be on circuits with the same architecture
at different nodes, allowing layout design difficulty to be compared
across nodes.
We will also evaluate more models and LLM-based layout generation
frameworks, with additional results reported on the
\href{https://lizhangmai.github.io/ICLayout-Bench/}{project website}.

\begingroup
\small
\raggedright
\setlength{\bibsep}{2pt}
\bibliographystyle{unsrtnat}
\bibliography{references}
\endgroup

\appendix
\crefalias{section}{appendix}
\section*{Appendix}
\numberwithin{table}{section}
\numberwithin{figure}{section}
\numberwithin{equation}{section}
\renewcommand{\theHtable}{appendix.\thesection.\arabic{table}}
\renewcommand{\theHfigure}{appendix.\thesection.\arabic{figure}}
\renewcommand{\theHequation}{appendix.\thesection.\arabic{equation}}

\section{Task Coverage}
\label{app:task-inventory}

\begingroup
  \small
  \setlength{\tabcolsep}{6pt}
  \renewcommand{\arraystretch}{1.22}
  \begin{xltabular}{\linewidth}{@{}P{2.5cm}Yrrrr@{}}
  \caption{Task counts for the full collections and evaluated subsets by
  (a) circuit category and (b) process. In (b), public tasks are grouped
  by PDK and private tasks by technology node.}
  \label{tab:inventory}\\
    \toprule
    & & \multicolumn{2}{c}{Public} & \multicolumn{2}{c}{Private} \\
    \cmidrule(lr){3-4}\cmidrule(l){5-6}
    \multicolumn{2}{@{}l}{Coverage}
    & Inventory & Evaluated & Inventory & Evaluated \\
    \midrule
    \endfirsthead
    \multicolumn{6}{@{}l}{Table \thetable{} (continued)}\\[3pt]
    \toprule
    & & \multicolumn{2}{c}{Public} & \multicolumn{2}{c}{Private} \\
    \cmidrule(lr){3-4}\cmidrule(l){5-6}
    \multicolumn{2}{@{}l}{Coverage}
    & Inventory & Evaluated & Inventory & Evaluated \\
    \midrule
    \endhead
    \bottomrule
    \endfoot
    \multicolumn{6}{@{}l}{\textit{(a) Circuit category}} \\
    Signal chain & Amplifiers & 58 & 10 & 7 & 4 \\
     & Clocks \& timing & 1 & 1 & 17 & 5 \\
     & Data converters & 2 & 1 & 3 & 1 \\
     & Interface & 1 & 1 & 2 & 1 \\
     & RF \& microwave & 1 & 0 & 0 & 0 \\
     & Sensors & 3 & 1 & 0 & 0 \\
     & Switches \& multiplexers & 5 & 0 & 4 & 2 \\
    \addlinespace[3pt]
    Power management & Linear \& LDO regulators & 8 & 2 & 2 & 1 \\
     & Voltage references & 1 & 0 & 3 & 1 \\
     & Current references \& sources & 1 & 0 & 2 & 0 \\
     & Supervisor \& reset ICs & 1 & 1 & 0 & 0 \\
    \addlinespace[3pt]
    Logic \& memory & Logic & 7 & 2 & 32 & 4 \\
     & Voltage translation & 0 & 0 & 1 & 0 \\
     & Memory & 3 & 2 & 0 & 0 \\
    \midrule
    \multicolumn{6}{@{}l}{\textit{(b) Process}} \\
    FreePDK45 & 45\,nm & 9 & 3 & --- & --- \\
    GF180MCU & 180\,nm & 27 & 6 & --- & --- \\
    IHP SG13G2 & 130\,nm & 49 & 8 & --- & --- \\
    SKY130 & 130\,nm & 7 & 4 & --- & --- \\
    \addlinespace[3pt]
    Private & 3\,nm & --- & --- & 6 & 1 \\
    Private & 12\,nm & --- & --- & 20 & 3 \\
    Private & 15\,nm & --- & --- & 4 & 2 \\
    Private & 28\,nm & --- & --- & 10 & 2 \\
    Private & 45\,nm & --- & --- & 1 & 1 \\
    Private & 90\,nm & --- & --- & 22 & 5 \\
    Private & 180\,nm & --- & --- & 10 & 5 \\
    \midrule
    \multicolumn{2}{@{}l}{Total tasks}
    & \benchPublicCount & 21
    & \benchPrivateCount & 19 \\
  \end{xltabular}
\endgroup

\section{Evaluation Validity and Scoring}
\label{app:interface}

\Cref{tab:appendix-normalization} specifies the quality functions used
to compute the credits in \Cref{sec:scoring}. Here $x$ is the post-layout
measurement and $a$ is its anchor; subscripts $i$ and $k$ identify the
metric and condition, respectively. The scale $s$ has the metric's unit;
target, degradation, and saturating forms require a positive scale.
Ratio objectives use nonnegative observations and positive denominators.
Each form satisfies $Q(a,a)=1$.
Each condition is normalized before the metric is aggregated. For the
fixed required conditions $\mathcal K_i$, the credited quality is
\begin{equation}
  c_i=\min\left(1,\min_{k\in\mathcal K_i}Q_i(x_{ik},a_{ik})\right).
  \label{eq:worst-quality}
\end{equation}
This order matters because source anchors can vary across conditions:
the worst raw measurement need not have the worst normalized quality.
The capped credits are used in \Cref{eq:score}, while raw measurements
still show improvements beyond the target.
In the case study, the electrical objectives are summarized separately by
$E=\prod_{i\in\mathcal E}c_i^{w_i/W_E}$, where $\mathcal E$ is the set
of weighted electrical objectives and $W_E=\sum_{i\in\mathcal E}w_i$.
This preserves their relative weights within the electrical component.

\begingroup
  \small
  \setlength{\tabcolsep}{4pt}
  \renewcommand{\arraystretch}{1.6}
  \begin{xltabular}{\linewidth}{@{}llY@{}}
  \caption{Electrical normalization functions $Q(x,a)$. Arrows indicate
  the preferred direction of the measured metric.}
  \label{tab:appendix-normalization}\\
    \toprule Form & $Q(x,a)$ & Use \\
    \midrule
    \endfirsthead
    \multicolumn{3}{@{}l}{Table \thetable{} (continued)}\\[3pt]
    \toprule Form & $Q(x,a)$ & Use \\
    \midrule
    \endhead
    \bottomrule
    \endfoot
    Ratio $\downarrow$ & $\dfrac{a+s}{x+s}$ & Lower-is-better metrics, such as delay. \\
    \addlinespace[3pt]
    Ratio $\uparrow$ & $\dfrac{x+s}{a+s}$ & Higher-is-better metrics, such as bandwidth. \\
    \addlinespace[3pt]
    Saturating ratio $\downarrow$ & $\dfrac{2(a+s)}{a+x+2s}$ & Nonnegative error with bounded improvement. \\
    \addlinespace[3pt]
    Decibel $\uparrow$ & $10^{(x-a)/20}$ & Higher-is-better amplitude metrics expressed in dB. \\
    \addlinespace[3pt]
    Decibel $\downarrow$ & $10^{(a-x)/20}$ & Lower-is-better amplitude metrics expressed in dB. \\
    \addlinespace[3pt]
    Target & $\dfrac{1}{1+|x-a|/s}$ & Preservation of an operating point. \\
    \addlinespace[3pt]
    Degradation $\downarrow$ & $\dfrac{s}{s+\max(0,x-a)}$ & Penalty for values above the anchor, scaled by $s$. \\
    \addlinespace[3pt]
    Degradation $\uparrow$ & $\dfrac{s}{s+\max(0,a-x)}$ & Penalty for values below the anchor, scaled by $s$. \\
  \end{xltabular}
\endgroup

\Needspace{0.95\textheight}
\section{Scores by Task}
\label{app:task-scores}

\begin{figure}[!htbp]
  \centering
  \includegraphics[width=\linewidth,height=18.8cm,keepaspectratio]{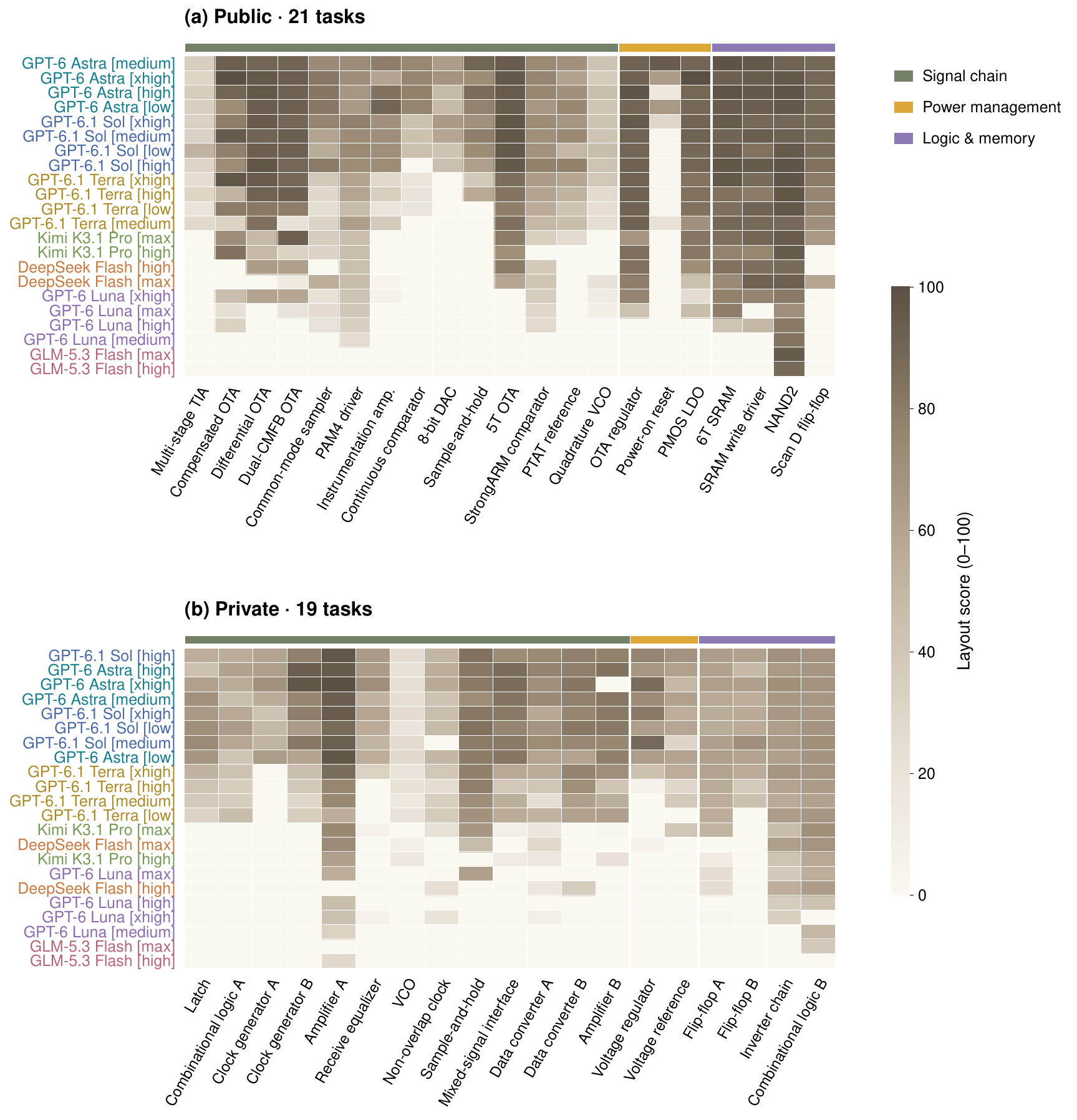}
  \caption{Per-task scores for the public (a) and private (b) subsets.
  Columns are ordered by circuit group and process; rows follow mean-score
  ranking. Shading encodes score (0--100).
  A and B distinguish separate circuits of the same function.}
  \label{fig:task-results}
\end{figure}

\Cref{fig:task-results} shows per-task scores for each configuration.

\end{document}